\documentclass[twocolumn]{aastex61}
\hypersetup{linkcolor=red,citecolor=blue,filecolor=cyan,urlcolor=magenta}
\submitjournal{ApJ}

\shorttitle{Prospects of determination of GR parameter $\beta$ with asteroid radar astronomy}
\shortauthors{Verma et al. 2017}
\begin{document}

\title{Prospects of dynamical determination of General Relativity parameter $\beta$ and solar quadrupole moment $J_{2\odot}$ with asteroid radar astronomy}

\correspondingauthor{Ashok Kumar Verma}
\email{ashokverma@ucla.edu}

\author[0000-0002-7209-0004]{Ashok K. Verma}
\affil{Department of Earth, Planetary, and Space Sciences, University of California, Los Angeles, CA 90095, USA}

\author[0000-0001-9798-1797]{Jean-Luc Margot}
\affiliation{Department of Earth, Planetary, and Space Sciences, University of California, Los Angeles, CA 90095, USA}
\affiliation{Department of Physics and Astronomy, University of California, Los Angeles, CA 90095, USA}

\author[0000-0001-8834-9423]{Adam H. Greenberg}
\affiliation{Department of Physics and Astronomy, University of California, Los Angeles, CA 90095, USA}



\begin{abstract}
We evaluated the prospects of quantifying the parameterized
post-Newtonian parameter $\beta$ and solar quadrupole moment $J_{2\odot}$
with observations of near-Earth asteroids with large orbital
precession rates (9 to 27 arcsec century$^{-1}$).  We considered existing
optical and radar astrometry, as well as radar astrometry that can
realistically be obtained with the Arecibo planetary radar in the next
five years.  Our sensitivity calculations relied on a traditional
covariance analysis and Monte Carlo simulations.  We found that
independent estimates of $\beta$ and $J_{2\odot}$ can be obtained with
precisions of $6\times10^{-4}$ and $3\times10^{-8}$,
respectively.  Because we assumed rather conservative observational
uncertainties, as is the usual practice when reporting radar
astrometry, it is likely that the actual precision will be closer to
$2\times10^{-4}$ and $10^{-8}$, respectively.  A purely dynamical
determination of solar oblateness with asteroid radar astronomy may
therefore rival the helioseismology determination.

\end{abstract}



%%%%%%%%%%%%%%%%%%%
\section{Introduction}
 \label{intro}
The parameterized post-Newtonian (PPN) formalism is a useful framework
for testing metric theories of gravity~\citep{will14}.  It consists of
10 dimensionless parameters that describe the general properties of
the metric.  In general relativity (GR), only 2 of the 10
parameters are non-zero. They are known as the
Eddington$-$Robertson$-$Schiff parameters $\gamma$ and $\beta$.  $\gamma$
represents the amount of curvature produced by a unit mass, and
$\beta$ represents the amount of nonlinearity in the superposition law
for gravity.

Several techniques have been used to place observational bounds on
these parameters~\citep{will14}, including observations of the bending
and delay of light by spacecraft tracking \citep[e.g.,]{Bertotti03} or
Very Long Baseline Interferometry \citep[e.g.,][]{Lambert09}, and
fitting of ephemerides to observations of planetary positions
\citep[e.g.,][]{Folkner09, Fienga11, Verma14, Fienga15}.

In GR, $\gamma$ and $\beta$ are equal to one.  Doppler tracking of the
$Cassini$ spacecraft has shown that $\gamma$ does not differ from one by
more than $2 \times 10^{-5}$ \citep{Bertotti03}.  Ephemeris-based
studies prior to 2009 indicated that $\beta - 1$ does not differ from
zero by more than 10$^{-4}$ \citep{Folkner09,Pitjeva14}.  More
recently, the availability of precise ranging data from the MESSENGER
Mercury orbiter~\citep{solo01} enabled improved
estimates~\citep{Verma14, Fienga15,Park17}.  Here, we evaluate the
prospect of asteroid orbit precession measurements to place more
stringent bounds on $\beta$.  We consider Earth-based radar
observations of near-Earth asteroids with perihelion shifts larger
than 10 arcsec century$^{-1}$.

Orbital precession can also be caused by the nonuniformity of the
gravity field that results from the oblate shape of the Sun.  The
solar oblateness is characterized by the solar quadrupole moment,
$J_{2\odot}$~\citep[e.g.,][]{kaul00}.  Simultaneous estimation of
$\beta$ and $J_{2\odot}$ requires that the precessional effects due to
GR and to the Sun's oblateness be disentangled.  Fortunately, GR is a
purely central effect, whereas the oblateness-induced precession has an
inclination dependence.  The two effects also have a different
distance dependence \citep{Misn73}.  As a result, observations of a
small sample of near-Earth asteroids with a variety of semi-major axes
and inclinations (Table \ref{target}) can in principle be used to
estimate $\beta$ and $J_{2\odot}$ \citep{Margot03,Margot09}.

Current estimates of the solar quadrupole moment are typically derived
on the basis of interior models of the Sun constrained by
helioseismology data \citep[e.g.,][]{Mecheri04, Antia08}.  The current
best value from the helioseismology literature is $J_{2\odot} = (2.2
\pm 0.1) \times 10^{-7}$ \citep{will14}.  Dynamical estimates that do
not rely on fits to helioseismology data yield similar values of
$J_{2\odot} = 2.3\pm 0.25 \times 10^{-7}$ \citep{Fienga15} and
$J_{2\odot} = 2.25\pm 0.09 \times 10^{-7}$ \citep{Park17}.
High-precision dynamical estimates are important to validate our
understanding of the interior structure of the Sun.

Our simulations of the determination of $\beta$ and $J_{2\odot}$ using a
variety of asteroid orbits suggest that independent values of $\beta$
and $J_{2\odot}$ can be obtained with satisfactory precision: with
the traditionally conservative assignment of radar uncertainties,
$\beta$ can be constrained at the $6\times10^{-4}$ level and
$J_{2\odot}$ can be constrained at the $3\times10^{-8}$ level.  With
uncertainties that more closely reflect measurement errors, this
precision may be improved by a factor of $\sim$3. (Section
\ref{simres}).

The outline of this paper is as follows.  In Section \ref{ca}, we
describe our choice of target asteroids.  In Section \ref{methods}, we
discuss the estimation of asteroid orbits with optical and radar
measurements.  Our dynamical model and data reduction procedures are
described in Section \ref{dynMod} and \ref{dataPro}, respectively.
Orbit determination results are presented in Section \ref{resOrb}.
Simulations of the determination of $\beta$ and $J_{2\odot}$ are described
in Section \ref{simres}.

%%%%%%%%%%%%%%%%%%%
\section {Target asteroids}
\label{ca}
The per-orbit secular advance in the angular position of the perihelion is given by
\citep{Misn73}
\begin{equation}
\label{raEq}
\delta\omega = \frac{6\pi GM_\odot}{a(1-e^2)c^2} \bigg[\frac{(2-\beta+2\gamma)}{3}\bigg] 
                   + \frac{6\pi}{2}R^2_\odot \frac{(1-3/2 \sin^2i)}{a^2(1-e^2)^2}J_{2\odot},
\end{equation}
where $\omega$ is the argument of perihelion, $GM_\odot$ is the Sun's
gravitational parameter, $R_\odot$ is the radius of the Sun, $c$ is
the speed of light, and $a$, $e$, and $i$ are the semi-major axis,
eccentricity, and orbital inclination (with respect to the solar
equator) of a planetary body, respectively.  Because both GR and solar
oblateness affect perihelion precession, estimates of $\beta$ and
$J_{2\odot}$ are highly correlated and it is desirable to track a
variety of solar system bodies with a range of $a$, $e$, $i$ values to
disentangle the two effects.

\begin{figure*}
\centering
\noindent
\includegraphics[width=35pc]{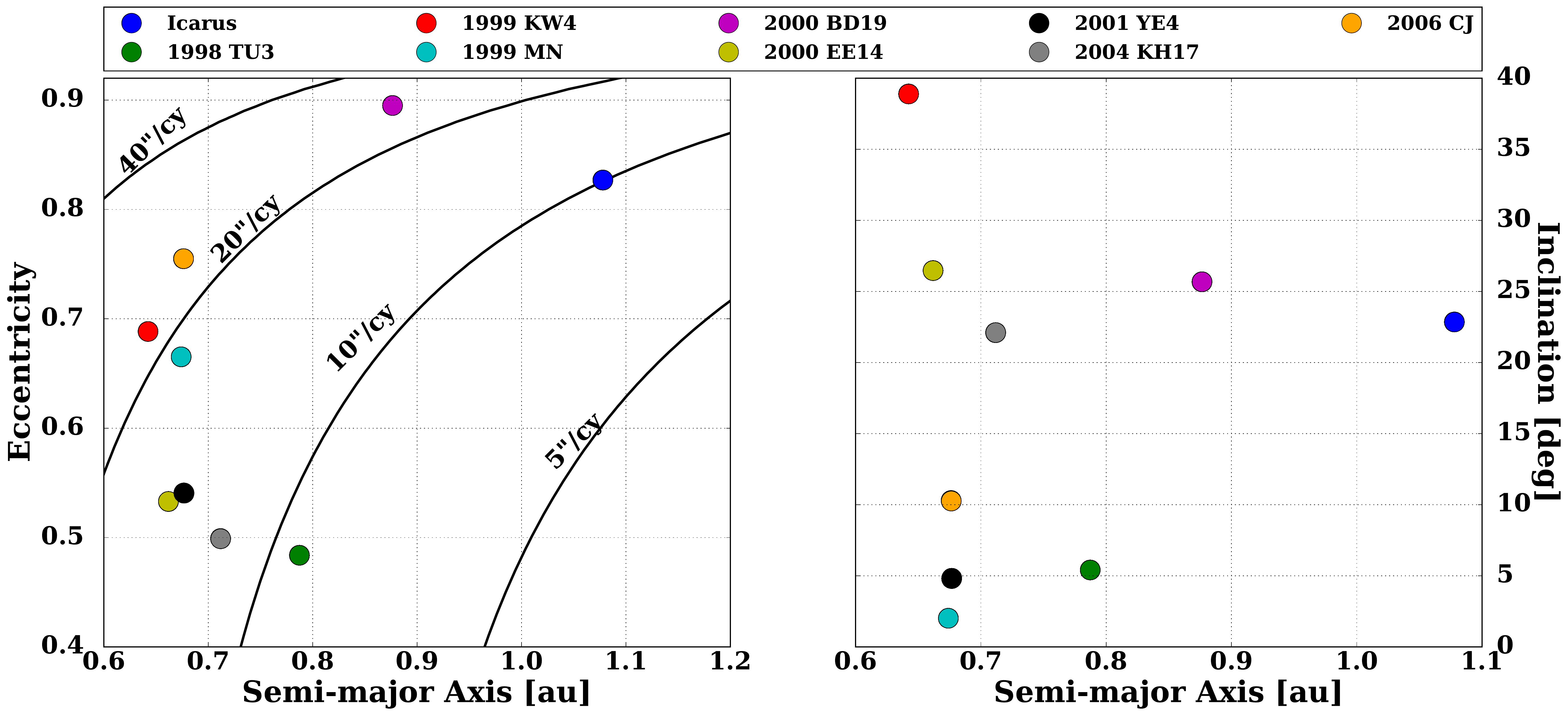}
\caption{Distribution of asteroid orbital elements for asteroids in our sample.  The corresponding rates of perihelion shift, predicted with Equation (\ref{raEq}), are shown as contour lines.}
\label{shift}  
\end{figure*}

Our selection of target asteroids follows the method of
\citet{Margot03}.  We select asteroids with both large perihelion
shift values and favorable observing conditions with radar (Table
\ref{target} and Figure \ref{shift}).  This sample of asteroid orbits
includes a wide range of semi-major axes, eccentricities, and
inclinations, which are advantageous when simultaneously solving for
$\beta$ and $J_{2\odot}$.  The predicted rates of perihelion advance,
$\dot{\delta\omega}$, shown in Figure \ref{shift} and Table
\ref{target} were computed assuming $\gamma = \beta = 1$ and
$J_{2\odot}=2.2 \times 10^{-7}$.

\begin{table}[h]
\footnotesize
\caption{Selected asteroids and orbital elements: Semimajor Axis ($a$), Eccentricity ($e$),
  and Inclination with Respect to the Ecliptic ($i_{\text{ec}}$) and Sun's equator ($i_{\text{eq}}$).}
 \centering
\begin{tabular}{ l r c c c c}
  \hline  
  \hline
    Target       &    $a$ (au) & $e$   &  $i_{\text{ec}}$ (deg)   & $i_{\text{eq}}$ (deg)   & $\dot{\delta\omega}$ ($''$ cy$^{-1}$)  \\ 
 \hline   
1566 Icarus  & 1.078  & 0.827 & 22.9   & 15.8   &  10.1  \\ 
 1998 TU3    & 0.787  & 0.484 &  5.41  &  3.41  &  9.11         \\     
 1999 KW4    & 0.642  & 0.688 & 38.9   &  46.0  &  22.1         \\
  1999 MN    & 0.674  & 0.665 &  2.02  &   5.25 &  18.5        \\  
2000 BD19    & 0.876  & 0.895 & 25.7   & 28.0   &  26.9          \\
2000 EE14    & 0.662  & 0.533 & 26.5   &  26.1  &  15.0          \\
 2001 YE4    & 0.677  & 0.541 &  4.82  &  11.0  &  14.4            \\
2004 KH17    & 0.712  & 0.499 & 22.1   & 14.9   &  12.0            \\
  2006 CJ    & 0.676  & 0.755 & 10.3   & 16.1   &  23.7            \\
\hline
\label{target}
\end{tabular}
\tablenotetext{}{{\bf{Note.}} The predicted rate of perihelion advance in arcsec century$^{-1}$
($''$ cy$^{-1}$), $\dot{\delta\omega}$, was computed using Equation (\ref{raEq}). }
\end{table}

%%%%%%%%%%%%%%%%%%%
\section {Methods}
\label{methods}

We first determined nominal trajectories for asteroids in our sample
with astrometric (i.e., positional) data, both optical and radar
(Table \ref{data}). The process involved three steps: (1) numerical
integration of each asteroid's orbit and calculation of partial
derivatives of the equations of motion with respect to the solve-for
parameters (i.e., the six components of the state vectors), (2) evaluation of 
simulated optical and radar observables
and computation of their partial derivatives with respect to the
solve-for parameters, and (3) least-squares adjustments to the solve-for 
parameters.

We used the Mission Operations and Navigation Toolkit Environment
(MONTE) software \citep[][MONTE v124]{Evans16}
for orbit determination and parameter estimation.
MONTE is an astrodynamics computing platform developed by
NASA's Jet Propulsion Laboratory (JPL). MONTE is used for spacecraft
navigation and trajectory design.  MONTE has also been used for a
variety of scientific purposes, including gravity analysis
\citep{Verma16} and ephemeris generation \citep{Greenberg17}.

\subsection{Dynamical model}
\label{dynMod}

MONTE uses a variable-step Adams-Bashforth method to numerically
integrate the equations of motion and corresponding partial
derivatives. Our dynamical model includes gravitational forces from
the Sun, 8 planets, and 21 minor planets with well-determined
masses \citep{Konopliv11}, general relativistic effects,
and perturbations due to the oblateness of the Sun.

In addition to these forces, we have also modeled the
nongravitational Yarkovsky orbital drift.  Perihelion advance due to
GR and solar oblateness does not affect the value of the semi-major
axis, but Yarkovsky drift does.  This nongravitational effect has
been shown to affect the semi-major axes of small bodies due to the
anisotropic re-emission of absorbed
sunlight~\citep[e.g.,][]{bott06areps}. The change in semi-major axis
with time due to Yarkovsky orbital drift, $\langle da/dt \rangle$, was
estimated for all target asteroids with the method of
\citet{Greenberg17}.  The values ranged in amplitude between 4 and 50
au/My, which is plausible for kilometer-sized bodies.  Only one target (1566
Icarus) is common between our target list and the 42 Yarkovsky detections
of \citet{nuge12yark}, and only one target (1999~MN) is common between our
target list and the 21 Yarkovsky detections of \citet{farn13yark}.  In
both cases, our Yarkovsky drift estimates are consistent with and
better constrained than prior work.

To initialize the integration process, we used a priori state vectors
extracted from the Minor Planet Center (MPC) database \citep{mpc}.

\subsection{Existing optical and radar astrometry}
\label{dataPro}
We used both optical and radar astrometry to determine the nominal
trajectory of each asteroid. Optical measurements provide positional
information on the plane of the sky.  They are typically expressed as
right ascension (R.A.) and declination (decl.) in the equatorial frame of
epoch J2000.0.  We downloaded optical astrometry from the MPC
\citep{mpc}.  We debiased optical astrometry and assigned data weights
according to the algorithm recommended by \cite{Farnocchia15}.

Radar astrometry consists of round-trip light time, a
measurement that can provide the asteroid$-$observer distance, and
Doppler shift, a measurement that can provide the line-of-sight
velocity of the asteroid with respect to the observer.
Radar measurements have fractional uncertainties as small as
10$^{-8}$.  The addition of radar astrometry can decrease orbital
element uncertainties by orders of magnitude compared to an
optical-only orbit solution~\citep{ostr04}.  However, the number of
radar measurements is typically small compared to the number of
optical observations (Table \ref{data}).

We processed a total of 12,102 optical measurements (R.A. and decl. pairs
obtained at 6051 epochs), as well as 56 range and 17 Doppler
measurements that have been published.

\subsection{Orbit determination for nominal trajectories}
\label{resOrb}
In order to compute nominal asteroid trajectories, we computed the
expected values of the observables and their partial derivatives with
respect to initial state vectors.  We calculated weighted residuals by
subtracting computed measurements ($C$) from observed measurements
($O$) and dividing the result by the corresponding observational
uncertainty ($\sigma$).  We adjusted initial state vectors with an
iterative least-squares techniques that minimized the sum of squares
of weighted residuals. Because there are 9 targets and 6 orbital
elements per asteroid in the nominal situation ($\gamma = 1,\ \beta =1,\ 
J_{2\odot} = 2.2 \times 10^{-7}$), we adjusted a total of 54
parameters.

We defined outliers as measurements with weighted
residuals in excess of three.  We identified and rejected 127 epochs with
outliers in the optical astrometry.  There were no outliers in the
radar astrometry.  We obtained a measure of the quality of the fit at
each iteration by computing the dimensionless rms of the weighted
residuals:
\begin{equation}
{\rm RMS} = \sqrt{\frac{1}{N}\sum_{i=1}^{N} \left( \frac{O_i-C_i}{\sigma_i} \right)^2},
\end{equation}
where $N$ is the number of observations, $O_i$ is the $i$th
observation, $C_i$ is the $i$th computed measurement, and
$\sigma_i$ is the observational uncertainty
associated with the $i$th observation.  We stopped the iterative
process when the change in the RMS of the weighted residuals between
two successive iterations was less than 0.01$\%$.
RMS residuals smaller than one indicate solutions that provide good
fits to the observations (Table \ref{data}).

\begin{table*}
\footnotesize
\caption{Selected asteroids and corresponding observations:
  Observational Interval, Number of Optical Pairs (R.A. and Decl.) of
  Observations, and Number of Published Range and Doppler
  Observations.}
 \centering
\begin{tabular*}{\textwidth}{l@{\extracolsep{\fill}}l c c c c c c c}
\hline  
\hline 
Target & Observational interval    & $N_{\rm opt}$   &  $N_{\rm rng}$ & $N_{\rm dop}$ & RMS$_{\rm opt}$ & RMS$_{\rm rng}$ & RMS$_{\rm dop}$ \\ 
 \hline   
1566 Icarus& 1949 Jun$-$2015 Jul& 1230 & 10 & 13& 0.56 & 0.28 & 1.10\\ 
 1998 TU3  &  1982 Dec$-$2016 Nov & 860  & $...$  & $...$ & 0.47 & $...$ & $...$\\     
 2000 BD19 &  1997 Feb$-$2016 Apr  & 522  & $...$  & $...$ & 0.51 & $...$ & $...$\\
 1999 KW4  &  1998 May$-$2016 Jul  & 2117 & 36 & $...$ & 0.39 & 0.39 & $...$\\
  1999 MN  &  1999 Jun$-$2015 Jun  & 141  & $...$  & $...$ & 0.64 & $...$ & $...$\\
2000 EE14  &  2000 Mar$-$2016 Jun  & 396  & $...$  & $...$ & 0.48 & $...$ & $...$\\
 2001 YE4  &  2001 Dec$-$2017 Jan  & 336  & 4  & 1 & 0.50 & 0.23 & 0.07\\
2004 KH17  &  2004 May$-$2016 May  & 211  & 1  & $...$ & 0.62 & 0.01 & $...$\\
  2006 CJ  &  2006 Feb$-$2017 Feb  & 238  & 5  & 3 & 0.59 & 0.30 & 0.11\\
\hline
\label{data}
\end{tabular*}
\tablenotetext{}{{\bf{Note.}} The last three columns provides the post-fit
  root-mean-square of weighted residuals.}
\end{table*}

\subsection{Anticipated radar astrometry}
\label{ss}
The objectives of this study are to evaluate the precision with which
PPN parameter $\beta$ and solar quadrupole moment $J_{2\odot}$ can be
determined from orbital fits constrained by existing and anticipated
optical and radar astrometry. To quantify the effect of anticipated
radar astrometry on the determination of these parameters, 
we simulated all existing optical and radar astrometry (Table \ref{data}) and
a number of anticipated Arecibo Observatory range measurements (Table \ref{simData})
with the nominal asteroid trajectories described above.
We did not attempt to simulate the effect of additional optical
astrometry, which is expected to improve the overall quality of the
fits, albeit not as powerfully as radar astrometry~\citep{ostr04}.

To supplement the published astrometry with realistic anticipated
values, we used the epochs of closest approach to Earth when the
asteroids are detectable with the Arecibo radar (Table \ref{simData}).
On the basis of prior experience, we assumed that two to four independent
data points would be collected at each future apparition.  For
apparitions in the past (identified in bold in Table \ref{simData}),
we used the number of data points that were actually obtained.  In
total, we simulated 61 independent range measurements in addition to
the 56 published values.  For each realization in our simulations, we
added noise to the observations by randomly drawing from a Gaussian
distribution with zero mean and standard deviation equal to the
observational uncertainty.  Observational uncertainties for
observations in the future were assigned according to signal-to-noise ratio
(S/N) and experience, with values ranging between 30 and 900 m.
Uncertainties for observations in the past mirrored the actual
measurement uncertainties adopted by the observer for these data
points.

\begin{table}[h]
\footnotesize
\caption{Selected asteroids and simulated observations: Years of Close
  Earth Approaches (yyyy $-$ 2000), Number of Simulated Radar Ranges, and
  Corresponding Uncertainties. }
\centering
\begin{tabular}{ l r c c}
\hline  
\hline 
    Target       &    Year of close approach & $N_{\rm range}$ & Uncertainties (m)\\ 
    \hline
 1998 TU3  &   $\bf{12}$, 19              & 5  & 75--900\\     
 1999 KW4  &  $\bf{16}$, 17, 18, 19, 20   & 12 & 40--300\\
  1999 MN  &   $\bf{04}$, $\bf{05}$       & 2  & 75--600\\
2000 BD19  &  $\bf{06}$, $\bf{07}$, 20    & 10 & 300--375\\
2000 EE14  & $\bf{07}$, $\bf{08}$, 21, 22 & 11 & 300--600\\
 2001 YE4  &  $\bf{12}$, $\bf{16}$, 21    & 10 & 30--600\\
2004 KH17  & $\bf{13}$                    & 2  & 300\\
  2006 CJ  &  $\bf{12}$, $\bf{17}$, 22    & 9  & 60--300\\
\hline
\label{simData}
\end{tabular}
\tablenotetext{}{{\bf{Note.}} Years
  highlighted in bold correspond to epochs for which data have already
  been collected.  The next detectable approach of 1566 Icarus is not until 2024.}
\end{table}

\subsection{Orbit determination with estimation of $\beta$ and $J_{2\odot}$}

We assigned solve-for parameters to one of two categories: local and
global.  Local parameters are specific to each asteroid, i.e., the 6
orbital elements or initial state vector (total of $9 \times 6 = 54$
parameters), whereas global parameters are common to all asteroids,
i.e., $\beta$ and $J_{2\odot}$.  We jointly solved for these 56
parameters.

We used two independent approaches to evaluate the precision in the
determination of global parameters $\beta$ and $J_{2\odot}$.  First, we
used a traditional covariance analysis (Section \ref{cova}) as
described in \citet{Bierman77}.
Second, we performed Monte Carlo simulations (Section \ref{full}) to
verify the results of the covariance analysis.

%%%%%%%%%%%%%%%%%%%
\section {Results}
\label{simres}

\subsection {Covariance analysis}
\label{cova}
A covariance analysis is a powerful technique that can be used to
evaluate the precision of solve-for parameters.  First, simulated,
noise-free measurements and their partial derivatives are computed on
the basis of nominal trajectories.  A least-squares estimation is then
performed, where the estimates logically converge on the nominal
values.  In the process, the associated covariance matrix is produced.
The expected precision of the estimated parameters is then inferred by
examining the covariance matrix. The square roots of the diagonal
elements provide the one-standard-deviation formal uncertainties.

After global  fits of 56 parameters, we obtained the following formal uncertainties:
\begin{equation}
  \sigma_{\beta} = 5.6 \times 10^{-4},
\end{equation}
\begin{equation}
\sigma_{J_{2\odot}} =  2.7 \times 10^{-8},
\end{equation}
with a correlation coefficient of -0.72.  The parameters remain
correlated because both GR and solar oblateness contribute to
perihelion precession.  However, the range of asteroid orbital
parameters (Table~\ref{target}) helps reduce the correlation
coefficient. Consideration of the Lense-Thirring effect for the Sun 
increases our $\sigma_{\beta}$ and $\sigma_{J_{2\odot}}$ estimates by 0.2\% and 4\%, respectively.

The expected formal uncertainty on $J_{2\odot}$ with direct dynamical
measurement of asteroids is 2.7 times the uncertainty based on
 fits to helioseismology data \citep{Antia08}.
For $\beta$, the expected formal uncertainty is
about twice the uncertainty obtained with pre-MESSENGER planetary
ephemerides~\citep{Konopliv11}, $\sim$7 times the uncertainty
obtained with post-MESSENGER planetary
ephemerides \citep{Verma14,will14,Fienga15}, and $\sim$14 times the uncertainty
obtained with MESSENGER range data \citep{Park17}
The formal uncertainties scale linearly with the uncertainties
assigned to the measurements.  It is often the case that radar
observers assign conservative uncertainties, as evidenced by RMS
residuals or reduced chi-square metrics that are almost always smaller
than unity and most often $<0.3$ (Table \ref{data}).  Therefore, we
anticipate that the actual precision may be improved by a factor of
$\sim$3, and the dynamical determination of $J_{2\odot}$ may be as
precise as the helioseismology determination.

In order to investigate the benefit of future observations, we also
performed covariance analyses under the assumption that observations
would stop at the end of 2017, 2019, or 2021, as opposed to 2022 in
our nominal scenario.  The results were
$\sigma_{\beta,2017} = 9.6  \times 10^{-4}$,
$\sigma_{\beta,2019} = 7.6  \times 10^{-4}$,
$\sigma_{\beta,2021} = 7.5  \times 10^{-4}$ and
$\sigma_{J_{2\odot},2017} = 1.9  \times 10^{-7}$,
$\sigma_{J_{2\odot},2019} = 4.2  \times 10^{-8}$,
$\sigma_{J_{2\odot},2021} = 3.8  \times 10^{-8}$.

\subsection {Monte Carlo simulations}
\label{full}

More robust results can be obtained by performing end-to-end
simulations that approximate the actual measurement and estimation
process.  In these analyses, integration of the trajectories and
estimation of the parameters are conducted as described in Section
\ref{methods} with two variations.  First, we chose initial values of
the solve-for parameters that are not identical to their nominal
values.  For instance, the initial positions and velocities of all
asteroids were changed by 10 km and 0.1 ms$^{-1}$ in each direction,
respectively.  Likewise, initial values for $\beta$ and $J_{2\odot}$
were changed  by $4 \times 10^{-4}$ and $5 \times 10^{-8}$, which is
approximately five times the uncertainty of recent estimates.
Second, we polluted the simulated measurements with independent noise
realizations as described in Section \ref{methods}.

We performed 500 Monte Carlo simulations.  
After convergence of the
least-squares estimation, we compared the estimated values of
solve-for parameters with their nominal values, which produced error
estimates.
To arrive at an estimate of the uncertainties, we can fit Gaussian
distributions to the histograms of error estimates, or we can compute
the covariance matrix, as follows:
\begin{equation}
\label{covEq}
\text{cov}(p_i, p_j) = \frac{1}{N-1}\sum_{k=1}^{N}(p_i^k - p_i^n)(p_j^k - p_j^n),
\end{equation}
where $N$ is the total number of simulations, $p_i^n$ is the nominal
value of the i$th$ parameter ($\beta=1$, $J_{2\odot} = 2.2
\times 10^{-7}$), and $p_i^k$ is the estimated value of
the i$th$ parameter from the k$th$ simulation of observations.
We used Equation (\ref{covEq}) and estimated the formal uncertainties in
the solve-for parameters by computing the square root of diagonal
elements.  We found
\begin{equation}
  \sigma_{\beta} = 7.4 \times 10^{-4},
\end{equation}
\begin{equation}
\sigma_{J_{2\odot}} =  3.7 \times 10^{-8}, 
\end{equation}
with a correlation coefficient of -0.81.  These values confirm the
covariance analysis results.

%%%%%%%%%%%%%%%%%%%
\section{Conclusions}
\label{conclusions}
A modest observing campaign requiring 50$-$60 hours of Arecibo telescope
time over the next five years can provide about 20 range measurements
of asteroids whose orbits exhibit large perihelion shift rates.  The
Arecibo Planetary Radar facility is required for these measurements
because its sensitivity is $\sim$20 times better than that of other
radar systems~\citep{naid16}, allowing detection of asteroids that are
not detectable elsewhere.

The Arecibo measurements will complement existing optical and radar
astrometry and enable joint orbital solutions with $\beta$ and
$J_{2\odot}$ as adjustable parameters.  Independent, purely dynamical
determinations of both parameters are important because they place
bounds on theories of gravity and the interior structure the of Sun,
respectively.

Our simulation results likely under-estimated actual precision for
two reasons. First, we did not attempt to simulate the impact of
future optical astrometry nor improvements to the accuracy of star
catalogs.  Both of these effects will inevitably improve the quality
of the orbital determinations.  Second, we assumed, based on
historical evidence, that radar observers assign fairly conservative
uncertainties to their measurements, which often underestimate the
precision of the measurements by a factor of $\sim$3 (Table~\ref{data}).  As a result, we
anticipate that the uncertainties of the final estimates will be close
to
\begin{equation}
  \sigma_{\beta} \sim 2 \times 10^{-4},
\end{equation}
\begin{equation}
\sigma_{J_{2\odot}} \sim 10^{-8}.
\end{equation}

%----------------------------------------------------------------------------------------
%	ACKNOWLEDGEMENTS
%----------------------------------------------------------------------------------------

\section*{Acknowledgments}

A.K.V., J.L.M., and A.H.G.\ were supported in part by the NASA Planetary
Astronomy program under grant NNX12AG34G.  J.L.M.\ and A.H.G.\ were supported
in part by NSF Planetary Astronomy program AST-0929830 and
AST-1109772.  This work was enabled in part by the Mission Operations
and Navigation Toolkit Environment (MONTE).  MONTE is developed at the
Jet Propulsion Laboratory, which is operated by Caltech under contract
with NASA.

\software{MONTE v124 \citep{Evans16}}

\bibliography{grsimul}

\begin{thebibliography}{}
\expandafter\ifx\csname natexlab\endcsname\relax\def\natexlab#1{#1}\fi
\providecommand{\url}[1]{\href{#1}{#1}}

\bibitem[{{Antia} {et~al.}(2008){Antia}, {Chitre}, \& {Gough}}]{Antia08}
{Antia}, H.~M., {Chitre}, S.~M., \& {Gough}, D.~O. 2008, A$\&$A, 477, 657

\bibitem[{{Bertotti} {et~al.}(2003){Bertotti}, {Iess}, \&
  {Tortora}}]{Bertotti03}
{Bertotti}, B., {Iess}, L., \& {Tortora}, P. 2003, nature, 425, 374

\bibitem[{{Bierman}(1977)}]{Bierman77}
{Bierman}, G.~J. 1977, {Factorization Methods for Discrete Sequential
  Estimation} (volume 128, 241 pages Academic Press, New York, NY, 1977)

\bibitem[{{Bottke} {et~al.}(2006){Bottke}, {Vokrouhlick{\'y}}, {Rubincam}, \&
  {Nesvorn{\'y}}}]{bott06areps}
{Bottke}, Jr., W.~F., {Vokrouhlick{\'y}}, D., {Rubincam}, D.~P., \&
  {Nesvorn{\'y}}, D. 2006, Annual Review of Earth and Planetary Sciences, 34,
  157

\bibitem[{{Evans} {et~al.}(2016){Evans}, {Taber}, {Drain}, {Smith}, {Wu},
  {Guevara}, {Sunseri}, \& {Evans}}]{Evans16}
{Evans}, S., {Taber}, W., {Drain}, T., {et~al.} 2016, in The 6th International
  Conference on Astrodynamics Tools and Techniques (ICATT), International
  Conference on Astrodynamics Tools and Techniques, Darmstadt, Germany.
\newblock
  \url{https://indico.esa.int/indico/event/111/session/30/contribution/177/material/paper/0.pdf}

\bibitem[{{Farnocchia} {et~al.}(2015){Farnocchia}, {Chesley}, {Chamberlin}, \&
  {Tholen}}]{Farnocchia15}
{Farnocchia}, D., {Chesley}, S.~R., {Chamberlin}, A.~B., \& {Tholen}, D.~J.
  2015, Icarus, 245, 94

\bibitem[{{Farnocchia} {et~al.}(2013){Farnocchia}, {Chesley},
  {Vokrouhlick{\'y}}, {Milani}, {Spoto}, \& {Bottke}}]{farn13yark}
{Farnocchia}, D., {Chesley}, S.~R., {Vokrouhlick{\'y}}, D., {et~al.} 2013,
  \icarus, 224, 1

\bibitem[{{Fienga} {et~al.}(2011){Fienga}, {Laskar}, {Kuchynka}, {Manche},
  {Desvignes}, {Gastineau}, {Cognard}, \& {Theureau}}]{Fienga11}
{Fienga}, A., {Laskar}, J., {Kuchynka}, P., {et~al.} 2011, Celestial Mechanics
  and Dynamical Astronomy, 111, 363

\bibitem[{{Fienga} {et~al.}(2015){Fienga}, {Manche}, {Laskar}, \&
  {Gastineau}}]{Fienga15}
{Fienga}, A., {Manche}, H., {Laskar}, J., \& {Gastineau}, M. 2015, Celest Mech
  Dyn Astr, 123:325

\bibitem[{{Folkner}(2009)}]{Folkner09}
{Folkner}, W.~M. 2009, in IAU Symposium, Vol. 261, IAU Symposium, ed.
  {S.~A.~Klioner, P.~K.~Seidelmann, \& M.~H.~Soffel}, 155--158

\bibitem[{{Greenberg} {et~al.}(2017){Greenberg}, {Margot}, {Verma}, {Taylor},
  {Naidu}, {Brozovic}, \& {Benner}}]{Greenberg17}
{Greenberg}, A.~H., {Margot}, J.-L., {Verma}, A.~K., {et~al.} 2017, \aj, 153,
  108

\bibitem[{{Kaula}(2000)}]{kaul00}
{Kaula}, W.~M. 2000, Theory of Satellite Geodesy: Applications of Satellites to
  Geodesy (Dover Publications, Mineola, NY), doi:10.1063/1.3033941

\bibitem[{{Konopliv} {et~al.}(2011){Konopliv}, {Asmar}, {Folkner}, {Karatekin},
  {Nunes}, {Smrekar}, {Yoder}, \& {Zuber}}]{Konopliv11}
{Konopliv}, A.~S., {Asmar}, S.~W., {Folkner}, W.~M., {et~al.} 2011, Icarus,
  211, 401

\bibitem[{{Lambert} \& {Le Poncin-Lafitte}(2009)}]{Lambert09}
{Lambert}, S.~B., \& {Le Poncin-Lafitte}, C. 2009, A$\&$A, 499, 331

\bibitem[{{Margot}(2003)}]{Margot03}
{Margot}, J.~L. 2003, in Bulletin of the American Astronomical Society,
  Vol.~35, AAS/Division of Dynamical Astronomy Meeting \#34, 1039

\bibitem[{{Margot} \& {Giorgini}(2009)}]{Margot09}
{Margot}, J.~L., \& {Giorgini}, J.~D. 2009, in IAU Symposium, Vol. 261, IAU
  Symposium, ed. {S.~A.~Klioner, P.~K.~Seidelmann, \& M.~H.~Soffel}, 183--188

\bibitem[{{Mecheri} {et~al.}(2004){Mecheri}, {Abdelatif}, {Irbah}, {Provost},
  \& {Berthomieu}}]{Mecheri04}
{Mecheri}, R., {Abdelatif}, T., {Irbah}, A., {Provost}, J., \& {Berthomieu}, G.
  2004, solphys, 222, 191

\bibitem[{{Minor Planet Center}(2017)}]{mpc}
{Minor Planet Center}. 2017, MPC database,
  http://www.minorplanetcenter.net/iau/mpc.html, ,

\bibitem[{{Misner} {et~al.}(1973){Misner}, {Thorne}, \& {Wheeler}}]{Misn73}
{Misner}, C.~W., {Thorne}, K.~S., \& {Wheeler}, J.~A. 1973, {Gravitation}

\bibitem[{{Naidu} {et~al.}(2016){Naidu}, {Benner}, {Margot}, {Busch}, \&
  {Taylor}}]{naid16}
{Naidu}, S.~P., {Benner}, L.~A.~M., {Margot}, J.~L., {Busch}, M.~W., \&
  {Taylor}, P.~A. 2016, \aj, 152, 99

\bibitem[{Nugent {et~al.}(2012)Nugent, Margot, Chesley, \&
  Vokrouhlick{\'y}}]{nuge12yark}
Nugent, C.~R., Margot, J.~L., Chesley, S.~R., \& Vokrouhlick{\'y}, D. 2012,
  Astronomical Journal, 144, 60

\bibitem[{{Ostro} \& {Giorgini}(2004)}]{ostr04}
{Ostro}, S.~J., \& {Giorgini}, J.~D. 2004, in Mitigation of Hazardous Comets
  and Asteroids, ed. M.~J.~S. {Belton}, T.~H. {Morgan}, N.~H. {Samarasinha}, \&
  D.~K. {Yeomans}, 38

\bibitem[{{Park} {et~al.}(2017){Park}, {Folkner}, {Konopliv}, {Williams},
  {Smith}, \& {Zuber}}]{Park17}
{Park}, R.~S., {Folkner}, W.~M., {Konopliv}, A.~S., {et~al.} 2017, \aj, 153,
  121

\bibitem[{{Pitjeva} \& {Pitjev}(2014)}]{Pitjeva14}
{Pitjeva}, E.~V., \& {Pitjev}, N.~P. 2014, Celestial Mechanics and Dynamical
  Astronomy, 119, 237

\bibitem[{{Solomon} {et~al.}(2001){Solomon}, {McNutt}, {Gold}, {Acu{\~n}a},
  {Baker}, {Boynton}, {Chapman}, {Cheng}, {Gloeckler}, {Head}, {Krimigis},
  {McClintock}, {Murchie}, {Peale}, {Phillips}, {Robinson}, {Slavin}, {Smith},
  {Strom}, {Trombka}, \& {Zuber}}]{solo01}
{Solomon}, S.~C., {McNutt}, R.~L., {Gold}, R.~E., {et~al.} 2001, \planss, 49,
  1445

\bibitem[{{Verma} {et~al.}(2014){Verma}, {Fienga}, {Laskar}, {Manche}, \&
  {Gastineau}}]{Verma14}
{Verma}, A.~K., {Fienga}, A., {Laskar}, J., {Manche}, H., \& {Gastineau}, M.
  2014, A$\&$A, 561, A115

\bibitem[{{Verma} \& {Margot}(2016)}]{Verma16}
{Verma}, A.~K., \& {Margot}, J.~L. 2016, Journal of Geophysical Research
  (Planets), 121, 1627

\bibitem[{{Will}(2014)}]{will14}
{Will}, C.~M. 2014, Living Reviews in Relativity, 17, 4

\end{thebibliography}



\end{document}